# Wearable Respiration Monitoring: Interpretable Inference with Context and Sensor Biomarkers

Ridwan Alam, *Student Member, IEEE*, David B. Peden, and John C. Lach, *Senior Member, IEEE*

*Abstract*—Breathing rate (BR), minute ventilation (VE), and other respiratory parameters are essential for real-time patient monitoring in many acute health conditions, such as asthma. The clinical standard for measuring respiration, namely Spirometry, is hardly suitable for continuous use. Wearables can track many physiological signals, like ECG and motion, yet not respiration. Deriving respiration from other modalities has become an area of active research. In this work, we infer respiratory parameters from wearable ECG and wrist motion signals. We propose a modular and generalizable classification-regression pipeline to utilize available context information, such as physical activity, in learning context-conditioned inference models. Morphological and power domain novel features from the wearable ECG are extracted to use with these models. Exploratory feature selection methods are incorporated in this pipeline to discover application-specific interpretable biomarkers. Using data from 15 subjects, we evaluate two implementations of the proposed pipeline: for inferring BR and VE. Each implementation compares generalized linear model, random forest, support vector machine, Gaussian process regression, and neighborhood component analysis as contextual regression models. Permutation, regularization, and relevance determination methods are used to rank the ECG features to identify robust ECG biomarkers across models and activities. This work demonstrates the potential of wearable sensors not only in continuous monitoring, but also in designing biomarker-driven preventive measures.

## I. Introduction

RESPIRATION tracking is vital for patients suffering from acute cardiopulmonary health conditions. Breathing rate (BR, also called respiratory rate), minute ventilation (VE, or minute volume), and other respiratory parameters are essential for assessing and forecasting risks of health crises such as cardiac arrest, sleep apnea, and asthma attack [1-5]. While these parameters are often incorporated into early warning and track-and-trigger systems at hospital wards [6], such preventive measures are still unavailable for patients at home due to the lack of continuous monitoring capabilities of respiratory parameters.

For example, the risk of exacerbation for asthmatic patients is often associated with short-term or sudden exposure to air pollutants, such as ozone ($O_3$), even indoor [7,8]. Personal exposure tracking strategies, based on the nociceptive lung function response to pollutants, require continuous monitoring of instantaneous minute ventilation, VE, which is the amount of air breathed in or out per minute. VE is a major factor in determining the "effective" dose of exposure; exposure to even moderate pollutant concentration level at high ventilation rate can induce complications in the lung function [9,10]. Hence, continuous VE monitoring can enable potential risk assessment to prevent exacerbation (Fig 1).

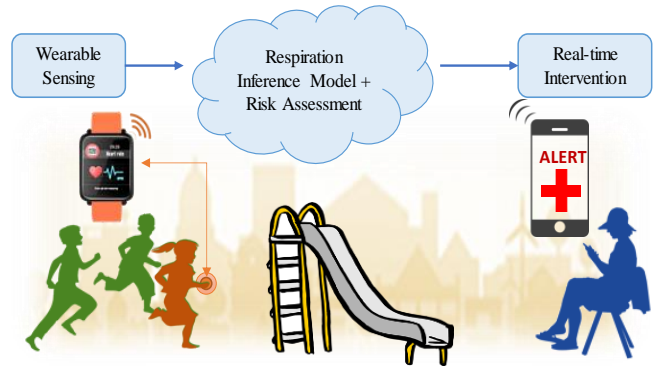

Fig. 1. Wearable-based continuous respiration monitoring for asthmatic children can reduce the risk of sudden exacerbation.

Spirometry is the clinically accepted standard for measuring respiratory parameters [11]. This modality, even in its portable form [12], is extremely invasive and not suitable for continuous day-to-day use. Hence, out-of-hospital or at-home continuous respiration monitoring remains an open challenge. Along with the direct measurement methods, indirect, or surrogate, measurements from other physiological signals, such as ECG-derived respiration (EDR), are gaining momentum. With the advent of wearable sensors, such methods can achieve the long-sought unobtrusiveness and usability. Yet, the challenge remains to improve the measurement performance against the noise and uncertainty in signals acquired using wearables [13-16].

With the motivation toward asthma attack prevention, we attempt to estimate the respiratory parameters, BR and VE, using wearable ECG and wrist-worn IMU sensors. Challenges toward this objective span from sensor noise reduction to physiological signal representation and modeling the relation-

This work was partly funded by National Institute of Environmental Health Sciences (R01-ES023349) and National Science Foundation Nanoscience Engineering Research Center for ASSIST (EEC-1160483).

R. Alam is a PhD student in Electrical Engineering at the University of Virginia, Charlottesville, VA 22904 USA (email: ridwan@virginia.edu).

D. Peden is the Director of the Center for Environmental Medicine, Asthma, and Lung Biology, University of North Carolina, Chapel Hill, NC 27599 USA (email: david_peden@med.unc.edu).

J. Lach is the Dean of Engineering and Professor at the Department of Electrical and Computer Engineering, George Washington University, Washington, DC 20052 USA (email: jlach@gwu.edu).

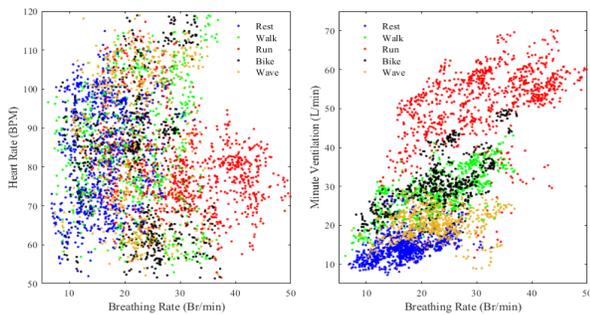

Fig. 2. Contextual variations in breathing rate against heart rate and against minute ventilation; data from 15 subjects ask for representative biomarkers and independent inference.

ship between sensor data and respiration. The inter-personal and contextual variations among the physiological parameters challenge the exploration of interpretable relationships (Fig. 2). Our earlier works explored the feasibility and utility of these sensor modalities for representation and prediction, reported the uncertainties in error distributions across physical activities, and asked for investigation into the contextual variations of the physiological relationships [17,18]. In this work, we explore the value of context in modeling respiration and understanding the modeled relationships. Context can often be scavenged from external sources including wearable sensing modalities. But, incorporating such context into the respiration estimation may yield specialized models lacking flexibility and generalization. Moreover, sensing-driven AI models often lack interpretability beyond model explanation, though such interpretation is highly sought for in many real-world applications [19,20].

In this work, we address these challenges by proposing a novel contextual inference pipeline. The pipeline hierarchically learns the context and the inference, and then aggregates those to predict response variables. It can also be used to perform feature selection based contextual biomarker discovery. We implement two pipelines for estimating BR and VE using state-of-the-art models, namely, generalized linear regression, random forest, support vector machine, Gaussian process regression, and neighborhood component analysis, and evaluate the generalizability, and robustness of these implementations. The novel contributions of this work are:
1) A hierarchical pipeline for context-aware inference applications, featuring modular and generalizable classification-regression layers with probabilistic aggregation;
2) A biomarker exploration methodology incorporated into the proposed pipeline to investigate the feature space for identification of interpretable physiological dynamics;
3) A set of wearable ECG features to represent the dynamics of the heart signals across contexts and individuals, as well as the associated cardio-respiratory functionalities;
4) Two implementations of the proposed pipeline to infer BR and VE using the proposed wearable ECG features, and to interpret the model predictions from physiological context.

This paper starts with a brief overview of existing works. We describe our study in Section III and propose the novel wearable ECG features in Section IV. The proposed inference pipeline and biomarker discovery method are presented in Sections V and VI. We discuss the result in Section VII, and conclude by mentioning the limitations and future plans.

## II. RELATED WORKS

Continuous respiration monitoring has been an active area of research for the last two decades yielding many disruptive technologies [13-16]. Recent research efforts are looking for non-invasiveness and day-to-day usability, either by designing body-worn, contactless sensing devices for direct measurement, or by estimating respiration from wearable sensing-driven non-respiratory signals, to achieve real-world applicability.

Direct respiration sensing methods try to capture any related physiological phenomena. For example, inductance plethysmography can track changes in thoraco-abdominal surface area during respiration using two transducer sinusoidal coils and an oscillator on the body [21,22]. Similarly, magnetometer plethysmography tracks changes in body volume by magnetometer transmitter-receivers [23]. Also, piezoresistive, piezoelectric, and capacitive sensors are explored to capture the respiration-time transthoracic modulation [24,25]. Non-contact modalities such as radar, optical, and thermal imaging have also been proposed to achieve contact-free respiration monitoring [14]. While such methods bring in the capabilities to measure respiratory parameters beyond breathing rate (BR), their performance and usability need evaluation beyond stationary, calibrated, location-specific lab settings across contextual and inter-person variability in free-living.

Recently, research efforts toward estimating respiration as surrogate or indirect measures from peripheral physiological sensors are gaining momentum, thanks to both the technological advances and the wide acceptances of wearable sensors. Modalities such as electrocardiogram (ECG) and photoplethysmogram (PPG) are at the center of these efforts trying to capture the physiological interaction between cardiac and respiratory functionalities. ECG derived respiration (EDR) methods often use signal processing techniques, such as power spectrum analysis, wavelet transform, empirical mode decomposition, to demodulate or extract the respiration signal, and then estimate the related parameters [26-32]. Such methods are often prone to propagation of reconstruction error, worsening estimation performance. These methods can provide some level of interpretability compared to other machine learning methods. ECG features are used with variants of principal component analysis in data-driven models to estimate respiratory parameters [33-35]. PPG-based methods follow similar processing and modeling techniques, while adding the benefit of non-invasiveness by acquiring the signal using wearables [35,36]. Most of these works focus on BR as a coupled parameter of heart rate, which often lack to overcome the contextual variations (Fig. 2). Other respiratory parameters, such as tidal volume and minute ventilation (VE), remain less investigated. A few works on VE estimation use multimodal sensing with ECG, PPG, and motion from inertial measurement units (IMU), where IMU signals are often used in denoising motion artefacts or extracting activity intensity [37-40]. However, the robustness of such methods against ambulatory noise from wearable modality, and against interpersonal and cross-context variations, is yet to be evaluated in real-world setting.

In this work, we try to achieve a robust, generalizable, and interpretable machine learning framework for modeling any respiratory parameters using wearable sensor-based physiological signals and contextual information.

## III. DATA COLLECTION

In this work, we explore the potential of wearable sensing, namely ECG and IMU, in tracking respiration continuously. This is part of a larger study, which aims to explore and quantify the risk of ozone-induced environmental asthma, and is supported by the National Institute of Environmental Health Sciences (R01-ES023349). This project pursues multiple experiments in parallel including the lung function response to ozone exposure and prospective interventions, and the personal and contextual variation in lung function. In this latter branch, we collect wearable sensor and respiration data using a physical exercise protocol, which are used in this work.

### A. Participants

15 healthy volunteers, 9 women and 6 men, participates in this study. The participants come from various ethnicities, and their health and fitness statuses are different. They may have mild asthmatic history, as this is not an exclusion criterion. The only exclusion criteria are pregnancy and/or tobacco use. Table 1 presents the demographic details of this population.

### B. Sensing Devices

Each participant wears two commercially available devices: a Shimmer3 ECG device on the chest and a Shimmer3 IMU device on the wrist. The ECG unit is programmed to collect three bipolar ECG channels (Leads-I, II, III). The IMU houses 3-axes accelerometer and 3-axes gyroscope sensors to capture the wrist motion. Both devices have on-board MSP430 microcontrollers that sample the ECG and the IMU signals. The ECG signals are sampled at 250 Hz with ADC gain adjusted to capture 800 mV differential range, and are stored on an on-board flash memory. IMU signals are also sampled at 250 Hz, and are recorded to on-board flash memory.

To acquire respiration measurements as ground truth during the data collection sessions, clinical Spirometers, comprising pneumotachometer (Hans Rudolph model #3830), amplifier (HR PA-1 series-1110), connector (series-7001), 2-way non-rebreathing Y-valve (series-2730), and data acquisition device, are used under human expert supervision. This device acquires BR in breaths per minute, inspire duration in seconds, tidal volume in liters, peak inspiratory flow in liters per seconds, and VE in liters per minutes.

### C. Activity Protocol

To capture the contextual variations of the lung functions, a multipart physical exercise protocol is designed. Each participant is assisted by an observer in following the protocol step-by-step. Before the experiment, the participant is instrumented with the wearable devices. The protocol uses a treadmill to facilitate some of the activities. The designed sequence consists of 3 walking and 2 running sessions on the treadmill, 2 stationary biking, 2 random hand-waving movements, and 3 rest periods (Fig. 3). The protocol allocates about three minutes for each activity, as well as a padding of two minutes of rest between consecutive activities. To allow the physiological changes related to an activity reach stable states, we do not collect respiration labels for that the first minute. After performing each activity for one minute, the participant is instrumented with the Spirometer mouthpiece and the nose clip. The participant resumes that activity for about two more minutes during which both respiration and wearable sensor data are acquired. Thus, each data collection session with all the five activities takes about 80 minutes, though both sensor data and respiration labels are acquired for about 20 minutes. This protocol is approved by the IRB of the University of North Carolina (UNC) at Chapel Hill. The sessions are conducted in a specialized physiology monitoring facility at the EPA Human Studies Facility in Chapel Hill, NC in partnership with the UNC Center for Environmental Medicine with a cooperative agreement (US EPA CR 83578501).

For each participant, continuous streams of raw sensor signals from the wearable modalities, IMU and ECG, during the activity protocol are acquired. These signals are processed and used to train the proposed pipeline with a goal to predict respiratory parameters, BR and VE.

TABLE I
DEMOGRAPHIC COMPOSITION

| Sex | Age (year) | Height (cm) | Weight (kg) | BMI (kg/m$^2$) | Race (White/All) |
|---|---|---|---|---|---|
| M | 25 ± 5.5 | 173 ± 4.8 | 77 ± 16.3 | 26 ± 5.8 | 2/6 |
| F | 22 ± 3.0 | 164 ± 7.7 | 67 ± 7.1 | 25 ± 4.3 | 5/9 |

## IV. FEATURE DESIGN

Our proposed respiration inference pipeline uses the two wearable sensing modalities, IMU and ECG, independently for two different learning tasks. We use the IMU data to learn the physical activity context, and the ECG signals to model the cardiorespiratory interaction in predicting BR and VE. Hence, our designs of the feature spaces for these two modalities are nearly independent. The sole dependence that is maintained between these processes is that of time synchronization, which is required for the final aggregation.

### A. Wrist Motion

IMU sensors capture the wrist motion by sensing associated directional force and rotation. We employ a 6-dof IMU on the left wrist of each participant. Both the accelerometer (x, y, z axes) and the gyroscope (x, y, z axes) are sampled at 250 Hz. The accelerometer can capture forces within ±4g range, and the gyroscope captures rotation velocity within ±360 degrees per second (dps) range.

#### 1) Windowing and Filtering

The six-dimensional IMU raw signals are windowed and preprocessed for feature extraction. A 15 seconds-long stream of the 6-d signals are used to extract a feature vector instance. We slide the windows with 80% overlap, which resulted in a feature vector every 3 seconds. The window length is heuristically chosen to capture low frequency changes in the signals.

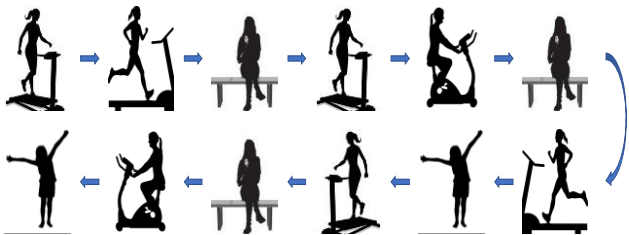

Fig. 3. Data is collected following a protocol that sequences five physical activities: walk, run, bike, wave, and rest.

Each windowed stream is filtered using a level clipper followed by a short-length median filter. This cascade addresses outliers and noises in the signals originating from communication, physical impact, and hardware issues. Moreover, to reduce effects of incidental motion artifacts, we apply a bandpass filter with a 0.01-20 Hz pass band.

*2) Feature Space*

We design the IMU feature space to contain standard statistical, frequency, and power domain features [41-43]. The statistical features are extracted as mean, max, median, standard deviation, rms, variance, and interquartile range of the raw IMU signal. Correlation and entropy are calculated pairwise of the accelerometer axes, and from those of the gyroscope, to capture spatial and rotational relationship. Frequency and power spectrum features are calculated in three frequency bands: 0.01-0.5 Hz, 0.5-3 Hz, and >3 Hz, along with the mean crossing rate. Teager energy operators are employed to extract mean, max, and variance of Teager energy of each signal. Thus, each 15-s signal window yields a feature instance of 90 features. And, every 3 seconds, a new feature instance is generated from the 6-d streams to be used by the context classifier.

### B. Wearable ECG

The chest-worn wearable device acquires ECG signal from three leads at 250 Hz sampling rate. To avoid the collinearity among the signals from those three leads, we use only Lead-I signal to acquire the features. The electrical activity of the heart makes the signal vary within about one millivolt range.

*1) Windowing and Filtering*

To capture stable patterns of the heart activity along with the dynamic variations across physical activities, we segment the raw wearable ECG signal using a window duration of 15 seconds. This window size is heuristically selected and may be varied across studies and sensing devices, if needed. But, to ensure temporal alignment of the wearable ECG feature instances with those from the wrist IMU modality, we ensure the feature extraction to be clocked every 3 seconds, which requires us to slide the window with 80% overlap.

The wearable ECG often suffers from disturbances due to baseline wondering, motion artefacts, and noise from skin contacts. Such noises are challenging and more prevalent in ambulatory and wearable ECG compared to stationary ECG. To reduce the effects of such disturbances, we first use median filtering to reduce speckle noises from skin contact or hardware issues. Then, we perform linear approximation of the baseline for each 15 s window and detrend the signal using that approximation. Finally, to reduce effects of motion artifacts, we use a bandpass filter with 5-25 Hz pass band on the detrended signal. This preprocessing stage improves the signal quality of all the windowed wearable ECG streams.

*2) Feature Space*

The wearable ECG feature space is designed to capture not only the overall characteristics of the heart's electrical activity within the time window, but also the dynamics among the individual beats within that window. Consequently, the feature space builds upon the morphological and frequency features extracted for a single beat. Our objective is to acquire the characteristics similar to those established for standard ECG beat [44,45]. For each preprocessed signal window, we imple-ment the standard peak detection algorithm to find the R-peak fiducial points, $\mathbf{r} = [r_1, r_2, …]$ using 200 ms lockout time. Since the signal within that window is detrended or baseline adjusted in the preprocessing step, we use a local threshold for the peak detector calculated as the 70% of the highest signal value. The temporal interval between consecutive peaks, i.e. the R-R intervals, are analyzed to identify possible missed peaks outside the ±20% deviation from the local average interval, and to update $\mathbf{r}$. Then, we search for Q, S, and T peak locations within the $r_i$-100 ms to $r_i$+500 ms segment for each peak in $\mathbf{r}$. Using these marks, we acquire the morphological characteristics, namely the magnitude, prominence, and width of the R-wave, the magnitude and width of the T-wave, the QS distance, and the ST distance of each heart beat (Fig. 4). We also calculate the beats-per-minute (BPM) from R-R interval, and the powers, defined as the area under the triangle, of the R-wave and the T-wave. Finally, we calculate the statistical mean and standard deviation of each of these features for the individual beats within the 15 s time window to acquire the corresponding 23-dimensional feature vector instance. To the best of our knowledge, this is a novel set of morphological and power features extracted from the wearable ECG using the standard signal processing algorithms for machine learning pipeline. Every 3 seconds, the wearable ECG feature extractor sends a feature instance to the regression inference models in our proposed pipeline to predict respiratory parameters.

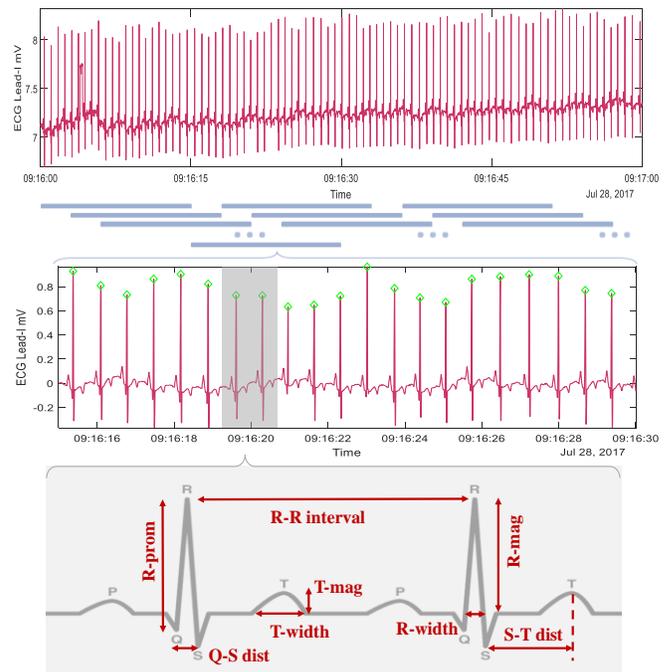

Fig. 4. Using sliding window segmentation before extracting morphological features from the wearable ECG streams.

### C. Respiration Response

The ground truth data for the respiratory parameters, BR and VE, are acquired asynchronously using the spirometer. We calculate the averages over 15 s window, same window length as motion and wearable ECG, to use as the response values for corresponding features. We also slide this window by 3 s, same as for the features, at each step to temporally sync the responses with the feature instances.

## V. Contextual Inference Pipeline

The wearable ECG and wrist motion features are used in a hierarchical classification-regression pipeline to infer the respiratory parameters (Fig. 5). There are three functionalities embedded in the pipeline: context classification, respiration regression, and contextual aggregation. We implement two instances of this pipeline for estimating two respiration parameters, BR and VE.

### A. Context Classifier

The context classifier is built to determine current physical activity context by using the wrist IMU features. This classifier can be trained independently of the regression models, hence, can be updated over time with possible changes in the context space or the predictor feature space. This modular design also enables the classifier to be transferable across applications. In our implementations, we train a single instance of the classifier and use that instance in both BR and VE inference pipelines.

We design the classifier as an ensemble of shallow decision trees using the totally corrective boosting algorithm, known as TotalBoost [46]. Unlike other boosting algorithms, this method updates the weight distribution for the "hard" examples in the training set by finding the distribution with minimum relative entropy to the initial distribution. In [46], this relative entropy is expressed as, $\Delta(d,d_0) = \sum_i d_i \ln(d_i/d_{0i})$, the KL divergence of two distributions. This algorithm prioritizes hypotheses that maximize the minimal margin of classification and minimize the number of observations below that margin, thus guarantees low generalization error [47].

Our implementation of TotalBoost uses shallow decision trees with maximum five splits. We design the decision trees to use Gini's diversity index as the metric for node splitting. We enforce choosing the split predictor based on chi-squared tests of independence not only between each predictor and the response, but also between each pair of predictors and the response. For the TotalBoost, we trained the ensemble with an upper bound of two hundred iterations. A margin precision parameter of $v = 0.01$ is used as a constraint in updating the hypothesis with respect to all past hypotheses. For an IMU feature instance $z_i$, the model yields the posterior probabilities $\mathbf{p}_i = [p^{(1)}, p^{(2)}, \ldots p^{(m)}, \ldots p^{(M)}]_i$, where M is the total number of contexts, and $p_i^{(m)} \in [0,1]$ is the posterior probability of that instance being in context m such that $\sum_m p_i^{(m)} = 1$.

### B. Respiration Regression Models

The pipeline incorporates a group of banks of regression models; each model learnt to infer contextual respiration from wearable ECG features. Each bank can be dedicated for one or more context, and is independent of other banks. A bank facilitates the modularity to use various context-specific models, depending on the application. The models within a bank are also independent of each other and operate in parallel (Fig. 5). For our implementations to infer BR and VE, we trained separate groups of banks of models. We explore five major categories of regression models for each contextual bank: generalized linear model, random forest, support vector machine, Gaussian process regression, and neighborhood component analysis. Each model in a context-bank is trained and operates independent of other models in that bank and those in other banks.

For the explanation of the model functionalities, let, each wearable ECG feature instance be represented by the row vector $\mathbf{x}_i = [x^{(1)}, x^{(2)}, \ldots x^{(d)}]_i$, as d is the number of features. Also, let the corresponding respiratory parameter be represented by the scalar $y_i$, for $i = 1, 2, \ldots n$; n is the size of the training set. In our two implementations, $y_i$ represents BR and VE, respectively. Each test instance feature vector is represented by $\mathbf{x}_t$ and the predicted respiration value as $\hat{y}$.

#### 1) Generalized Linear Model (GLM)

GLM extends linear regression by allowing for exponential distributions of the prediction error. While linear regression models the response variables to linearly vary with predictors, in GLM, a link function of the distribution mean of the response is expected to vary linearly with predictors [48]. Assuming an exponential distribution for the response, the link function f of the distribution mean μ is modeled against the feature instances $\mathbf{x}_i$'s using coefficient set $\beta = [\beta_0, \beta_1, \ldots \beta_d]$. This is formulated as $E[y] = \mu = f^{-1}(\beta \mathbf{X})$. The parameter β can be constrained using the elastic net with regularization parameter λ and scaling factor α. Thus, elastic net drives some coefficients to zero and reduce dimensionality [49,50], by minimizing the cost function, L(β), defined using the deviance of the model fit:

$$L(\beta) = \frac{1}{n}\text{Deviance}(\beta) + \lambda \frac{(1-\alpha)}{2}\|\beta\|_2^2 + \lambda\alpha\|\beta\|_1 \quad (1)$$

Our implementation of GLM uses the identity function as the link function f, assuming normal distribution for the respiration parameters, BR and VE. The coefficient β is learnt from the ECG features to model the distribution mean of the respiratory parameters. We dynamically adjust the regularization by calculating λ from the training sample size for each context bank. We combine both $L^1$ and $L^2$ penalties on β using $\alpha = 0.5$.

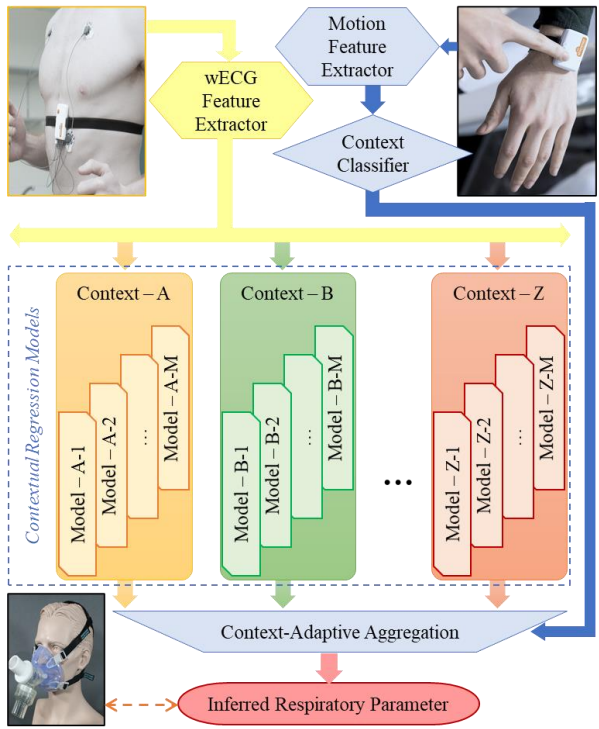

Fig. 5. Proposed classification-regression pipeline: Context classified from wrist motion is used to aggregate respiratory parameters independently inferred from wearable ECG.

### 2) Random Forest (RF)

RF regression aggregates the predictions from the ensemble of deep decision trees, each trained using N out of N instances randomly sampled with substitution from the training set. Each trained tree uses random subset of the predictors for splitting each node to avoid correlated trees in the ensemble [51]. The individual inferences of these weak learners are averaged to get the prediction from the ensemble. Permuting one predictor at a time, out-of-bag losses are analyzed to rank the predictors based on their contribution on the prediction [52].

For each context-bank, we employ an ensemble of two hundred decision trees to learn the respiratory parameter from the ECG feature set. These trained trees are designed to grow deep, while preventing overfitting to possible outliers by enforcing at-least ten observations at the leaf nodes. The mean squared error is used as the split criterion for these regression trees. To avoid the bias in the predictor selection at each node split, we address the interactions between the predictors by using interaction test. This method conducts chi-square tests of independence between each predictor and the response, as well as between each pair of predictors and the response. We prioritize the predictor that minimizes the p-values for both tests.

### 3) Support Vector Machine (SVM)

SVM regression uses a kernel-based transformation of the feature space, and learns an optimal hyperplane that limits the prediction error within an "insensitivity" threshold ε. The hyperplane is characterized by the support vectors, and is learnt as the coefficients $\alpha = [\alpha_1, \alpha_2, \ldots \alpha_n]$ and bias b by minimizing the loss function, $L(\alpha)$, defined in [53] as:

$$L(\alpha) = \frac{1}{2}\sum_{i,j=1}^{n}(\alpha_i - \alpha_i^*)(\alpha_j - \alpha_j^*)G(\mathbf{x}_i, \mathbf{x}_j) + \varepsilon\sum_{i=1}^{n}(\alpha_i + \alpha_i^*) - \sum_{i=1}^{n}y_i(\alpha_i - \alpha_i^*)$$

under constraints on $\alpha_i$'s using the box constraint C. The optimization process is also constrained by the Karush-Kuhn-Tucker complementarity conditions [53]. Using a Gaussian kernel for the transformation, a new feature instance is used to predict the corresponding response as:

$$\hat{y} = f(\mathbf{x}_t) = \sum_i (\alpha_i - \alpha_i^*) G(\mathbf{x}_i, \mathbf{x}_t) + b; \ G(\mathbf{x}_j, \mathbf{x}_k) = e^{-\|\mathbf{x}_j - \mathbf{x}_k\|^2}$$

Our implementations of the SVM regression dynamically adjusts the value of ε based on the distribution of the response variables. The box constraints are similarly adjusted as ten times that of ε. We use the sequential minimal optimization as the algorithm for minimizing the cost function with $10^{-6}$ feasibility gap as the associated convergence criterion.

### 4) Gaussian Process Regression (GPR)

GPR is a non-parametric kernel-based approach. In this probabilistic method, the response, $y_i$, is explained using a latent function of the predictors, $f(x_i)$, along with the linear combination of a transformation $h(x_i)$ of the predictor space [54]:

$$P(y_i|f(\mathbf{x}_i), \mathbf{x}_i) \sim \mathcal{N}(h(\mathbf{x}_i)^T\beta + f(\mathbf{x}_i), \sigma^2) \quad (2)$$

Here, the basis function h is a transformation of the feature space, chosen empirically. The linear combination coefficient vector β, the latent function f, and the noise variance, $\sigma^2$ is learnt from the data. The latent variables, $f_i = f(\mathbf{x}_i)$, are assumed to possess a Gaussian process prior, such that for all variables, $\mathbf{f} = [f_1, f_2, \ldots f_n]$, we get $P(\mathbf{f}|\mathbf{x}_1, \mathbf{x}_2, \ldots \mathbf{x}_n) \sim \mathcal{N}(0, K)$. For this prior, K is the covariance matrix defined using the kernel function k, as $K_{ij} = k(\mathbf{x}_i, \mathbf{x}_j)$. The parameter, θ, associated with the choice of the kernel function is learnt during training. Using estimated β, θ, and $\sigma^2$, the latent variable $\hat{f} = f(\mathbf{x}_t)$ is inferred for any test instance $\mathbf{x}_t$. The joint GP prior $P(\hat{f}, \mathbf{f})$ is used with the likelihood for $\mathbf{y} = [y_1, y_2, \ldots y_n]$, which is $P(\mathbf{y}|\mathbf{f})$, to get the joint posterior:

$$P(\hat{f}, \mathbf{f}|\mathbf{y}) = \frac{P(\hat{f}, \mathbf{f})P(\mathbf{y}|\mathbf{f})}{P(\mathbf{y})}$$

where $P(\hat{f}, \mathbf{f}) \sim \mathcal{N}\left(0, \begin{bmatrix} K_{\mathbf{f},\mathbf{f}} & K_{\hat{f},\mathbf{f}} \\ K_{\mathbf{f},\hat{f}} & K_{\hat{f},\hat{f}} \end{bmatrix}\right)$ and $P(\mathbf{y}|\mathbf{f}) \sim \mathcal{N}(\mathbf{h}^T\beta + \mathbf{f}, \sigma^2 I)$.

We marginalize this posterior over $\mathbf{f}$ to acquire $\hat{f}$, which is used as in (2) to get the response $\hat{y}$, i.e. the respiratory parameters.

Depending on the kernel function k, the covariance matrix K captures the similarity among feature instances. Parameters of these kernel functions are the signal variance, $\sigma_s^2$, and the characteristic length scale, $\sigma_l^2$. Automatic relevance determination (ARD) uses different length scale parameter $\sigma_r^2$ for each feature r = 1, 2, … d, to investigate their individual contribution in inferring the latent and the response variables [55].

We implement GPR and ARD by choosing the Matern kernel function k with separate $\sigma_r^2$ for each feature, defined as:

$$k(\mathbf{x}_i, \mathbf{x}_j) = \sigma_s^2 (1 + \sqrt{3m}) \exp(-\sqrt{3m}); \ m = \sum_{r=1}^{d} \frac{(x_i^{(r)} - x_j^{(r)})^2}{\sigma_r^2} \quad (3)$$

Here, the parameters, $\theta = [\sigma_s^2, \sigma_r^2]$, are learnt during training. To avoid local minima, we initialize $\sigma_s^2$ using the variance of the response variable, and $\sigma_r^2$ using feature variances. The noise variance $\sigma^2$ is initialized similarly as $\sigma_s^2$. For transforming the feature space, we use linear basis function $h(\mathbf{x}_i) = [1 \ \mathbf{x}_i]$, and learn the coefficients β from the data.

### 5) Neighborhood Component Analysis (NCA)

NCA is non-parametric and avoids any assumption about the sample distribution. It uses a stochastic neighbor selection rule to assign any test instance the response value of its selected neighbor. This rule reduces its dependence on the amount of training data and the risk of overfitting. NCA attempts to learn a quadratic distance metric, representable as linear transformation to low-dimensional input space, minimizing the regression loss [56,57]. For any $\mathbf{x}_s$ in training set S and a test instance $\mathbf{x}_t$, distance metric $D_w$ is defined using predictor weights, $w_r$, as,

$$D_w(\mathbf{x}_t, \mathbf{x}_s) = \sum_{r=1}^{d} w_r^2 |x_{tr} - x_{sr}| \quad (4)$$

Then, the stochastic selection uses the probability of any $\mathbf{x}_s$ in S being the nearest neighbor of $\mathbf{x}_t$ as $p_{ts}$:

$$p_{ts} = P(neigh(\mathbf{x}_t) = \mathbf{x}_s|S) = \frac{\exp(-\|D_w(\mathbf{x}_t, \mathbf{x}_s)\|)}{\sum_{s \in S} \exp(-\|D_w(\mathbf{x}_t, \mathbf{x}_s)\|)}$$

Using the response of the nearest neighbor relative to the learnt distance metric, the test response $\hat{y}$ is inferred.

Our implementation uses mean absolute error as the metric for measuring the regression loss, and learns the distance metric using the limited memory Broyden-Fletcher-Goldfarb-Shanno algorithm. Instead of storing all the training samples, we store the linear transformations. For each context-bank, we dynamically choose the regularization parameter based on the size of the training set for that bank.

### C. Context-Conditioned Aggregator

The final step of the pipeline is the aggregator that combines the inferences from the contextual regression banks based on the output of the context classifier. Traditional regression aggregators attempt either to select the best performing regressor from a group or to average their performances to achieve overall better performance [58]. Our proposed pipeline incorporates a novel conditional aggregation method merging both strategies depending on the context. Based on the level of the posterior probability for the classified context of an instance, we either select the regression bank, or use weighted averaging of the regression inferences from the contextual banks. The posterior probabilities for each context are used as the averaging weights.

Our two implementations of this pipeline, for BR and VE, are trained to use IMU motion and wearable ECG signals for predicting the respiratory parameters.

## VI. ECG Biomarker Discovery

Beyond the respiration inference, the proposed pipeline also facilitates sensor biomarker discovery from wearable signals. For health applications, the objective of such approach is two folds: first, to identify some tangible parameters that possess some 'meaning' within context, and second, to use some metric of those parameters for explaining the underlying relationship. Existing approaches for such exploration root from standard feature selection methods [19,20,43]. The feature ranks from these methods vary across models as the features are evaluated by their relevance to the predictions. This approach succeeds in explaining the mechanism of learnt models, but lacks physiological interpretation. Our approach builds on such techniques and explores interpretability by incorporating physiological and contextual perspectives into merging different model outcomes. In our BR and VE pipelines, we identify some wearable ECG biomarkers along with their contextual relevance, and interpret in light of related physiological functions. This is done in two steps: first, we rank the features using model specific metrics, and then, we selectively merge the features and their relevance from physiological and contextual perspectives.

### A. Regression Feature Ranking

We adapt the respiration regression model banks, presented in Section V, to acquire feature importance and relevance using model specific metrics.

#### 1) GLM with Elastic Net

Our GLM implementation uses the elastic net to regularize the $L^1$ and $L^2$ norms of the linear coefficients $\beta$, as shown in (1). In linearly combining the features to acquire the mean of the response distribution, smaller coefficients refer to smaller projections along those feature dimensions indicating less correlation or dependence. Regularization in (1) drives such smaller coefficients toward zero [50]. Using the learnt $\beta$, we calculate the feature weights representing the respective dependence of the link function of the distribution mean on these features.

#### 2) RF with Out-of-Bag Permutation

Using the out-of-bag instances of each learnt tree in RF, we randomly permute one of the features at a time over those instances and evaluate the effect on the inferred response values by measuring the out-of-bag losses. For important features, such permutation is assumed to affect the inference more [52]. Based on this assumption, we compare the out-of-bag losses between with and without permutation for each feature and use their differences to rank the features.

#### 3) Relevance Determination in GPR

In GPR, we use separate length scale parameters, $\sigma_r^2$, for each feature in defining the kernel function in (3), for $r = 1, 2, \ldots d$. During training, these parameters are learnt to build the kernel matrix K for the latent variables **f**. Low value for $\sigma_r^2$ represents high influence of the corresponding feature, as this low variance yields high value for the covariance function k [55]. We use these learnt parameters to evaluate the relevance and importance of each of the features, as $\exp(-\sigma_r^2)$, driving the relevance to zero for high length scales, and acquire the feature ranks.

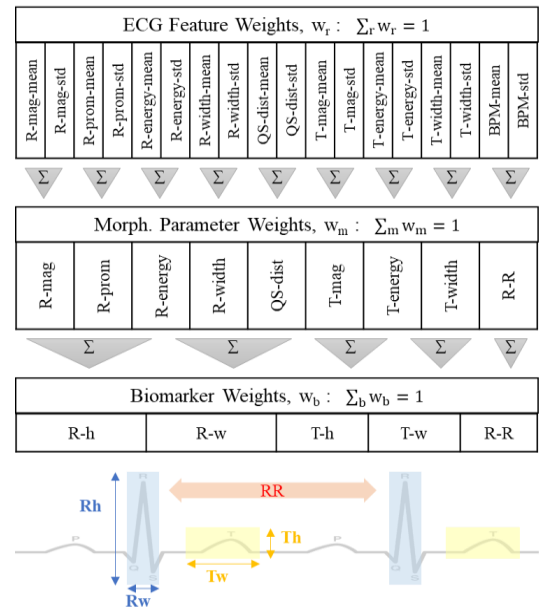

Fig. 6. Wearable ECG feature space is clustered to identify interpretable biomarkers; the feature weights are aggregated accordingly to acquire contextual relevance of biomarkers.

#### 4) NCA Distance Metric

For NCA, the feature weights, $w_r$, $r = 1, 2, \ldots d$, are incorporated as parameters for the distance metric, as shown in (4). During training, we optimize the inference loss, which depends on the stochastic neighbor selection rule defined using the distance metric, along with regularization on the norm of $w_r$. Such regularization drives the $w_r$ to zero for some features, reducing the dependence between the learnt distance metric and those features in characterizing the neighborhoods and their associated similarity in response variables [57]. We use these feature weights to rank the features accordingly.

### B. Interpretable Aggregation

Each context-bank of respiration regression models yields a set of feature weight vectors, $w_r$; elements of that vector represent the relevance of corresponding features, for that model in that bank. These weights vary across models, contexts, and the inferred respiratory parameters (BR and VE). We selectively cluster those features and their weights to achieve an aggregate of 'meaningful' variables from a physiological perspective.

As presented in Section IV, the wearable ECG feature space is designed using physiological knowledge-based signal processing methods. Hence, we start our clustering approach by first reverse mapping the feature space to associated electro-cardiac signal space (Fig. 6). We merge the statistical branches, namely the statistical mean and the standard deviation, associated with each ECG morphological parameter; for example, adding the weights for R-mag-mean and R-mag-std to get the relevance for R-mag. This step generates a set of lower dimensional weight vectors, $w_m$. This representation shows the contributions of individual ECG parameters, yet remain prone to the inherent dependence among these parameters, and consequent variations across models and contexts. To remove such dependence, we further cluster these parameters utilizing the morphological and functional knowledge. Parameters that are parallel in the 2-d ECG plane and originate from same electro-cardiac functions, for example R-width and QS-distance, we cluster those as a biomarker and acquire its relevance, $w_b$, by averaging the weights of corresponding parameters in $w_m$. The resulting set of biomarkers and the relevance vector contain only five elements each; R-wave height (Rh), R-wave width (Rw), T-wave height (Th), T-wave width (Tw), and R-R distance (RR). Rh and Rw refer, respectively, to the intensity and duration of the ventricular depolarization of the heart, Th and Tw similarly refer to the ventricular repolarization, and RR refers to the heart rhythm of this electro-cardiac function [44]. Because of the functional correspondence, these biomarkers can add interpretability about the physiological relationship, between the cardiac functions and the respiration, captured by the models.

We analyze the identified biomarkers and their relevance to individual model predictions in contextual perspective utilizing the inference pipeline. For any context, we compare and merge the biomarker relevance by averaging over the models in that context bank. The relevance values correspond to the impact of biomarkers on a model prediction in a certain context, and the associated cardio-respiratory relationship for that context.

## VII. RESULTS & DISCUSSION

We collected data from 15 healthy subjects, each performing a physical exercise protocol yielding about 15 minutes of sensor and respiration data for the five physical activities: rest, walk, run, bike, and wave (excluding rest periods between activities). The preprocessing generates 16 instances per minute, totaling about 3450 samples of data. Using these data, we evaluate the proposed pipeline for context classification, respiration inference, and biomarker relevance for BR and VE inference. To demonstrate the generalizability of the implemented methods, we conduct performance evaluation over a range of train-test hold-out percentages from 80% training - 20% testing to 70-30, 60-40, 50-50, 40-60, 30-70, and 20-80 percentages.

### A. Context Classification

For both BR and VE inference pipelines, we use the same context classifier that identifies the physical activities from the IMU motion sensor-based features. For each train-test ratio, we use the hold-out test set to evaluate the trained classifier with metrics such as accuracy, true positive rate (TPR), and false negative rate (FNR). The resulting scores and the confusion matrices for four ratios are presented in Fig. 7. Over all ratios, the mean accuracy is 99.66% with a range from 99.5% to 99.9%. For any context across the train-test ratios, the lowest TPR is 98.4% and the highest FPR is 1.6%. This result shows the robustness and generalizability of the classifier, even when trained on only 20% and tested on the rest of the data.

### B. Respiration Inference

Using the ECG features and the contextual pipeline, we infer two respiratory parameters, BR and VE. The inference loss is evaluated using mean absolute error (MAE) as shown in Fig. 8. For BR, this loss is calculated in breaths per minute (Br/min), and VE in liters per minute (L/min). Fig. 8 shows the results for the evaluation with 70% training and 30% hold-out test data. For this evaluation, the best performance, for both BR and VE, is acquired with NCA as the contextual regression model in the implemented pipelines. Here, for BR inference, the mean loss over all activities is 1.17 Br/min; including 0.7 Br/min during rest to 1.39 Br/min during run. And, for VE, the overall loss is 1.39 L/min, with 0.87 L/min at rest and 1.87 L/min during run. Similarly, overall losses for using GPR contextual models are, respectively, 1.32 Br/min and 1.46 L/min; and, for using SVM, 1.59 Br/min and 1.75 L/min. These losses are notably lower, for similar contexts, compared to existing solutions, even with stationary ECG, as well as our earlier works [17,28,36]. This result demonstrates the value of the novel wearable ECG features in capturing the physiological relationship.

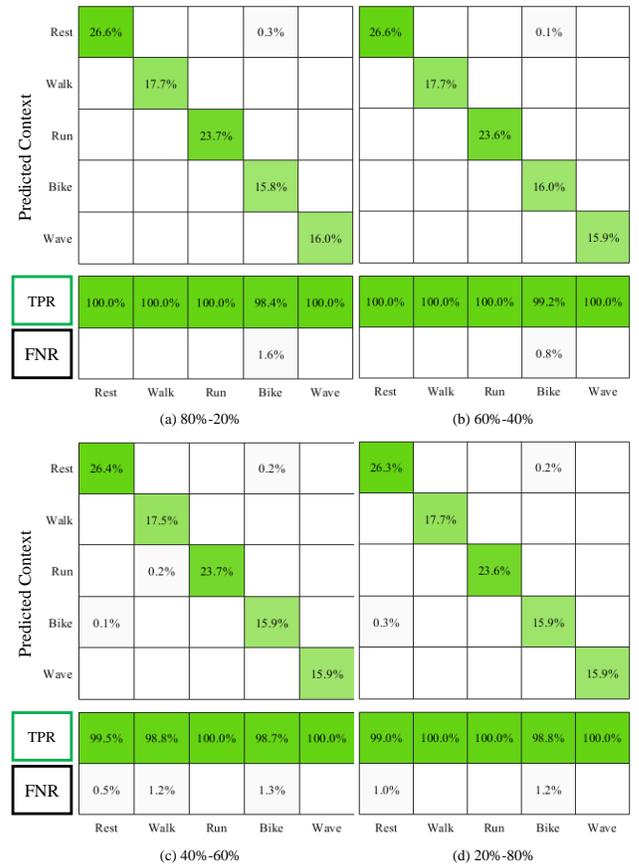

Fig. 7. Context classification performance over varying train-test hold-out percentages: training with (a) 80% data, (b) 60%, (c) 40%, and (d) 20% only, yet maintaining ≥99.5% accuracy.

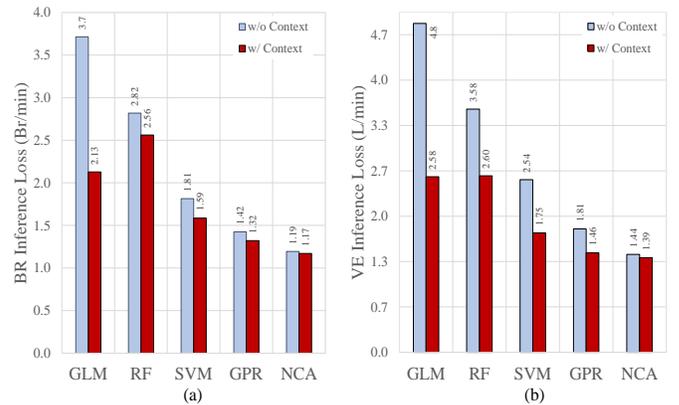

Fig. 8. MAE inference loss of the proposed contextual pipeline with different regression models and context-agnostic models of same kind: (a) breathing rate and (b) minute ventilation.

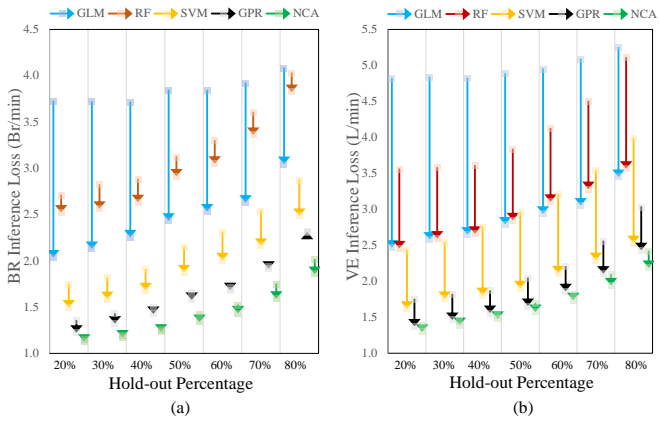

Fig. 9. Changes (magnitude and direction, shown as arrows) in inference loss from context-agnostic to proposed contextual models across varying train-test hold-out percentages.

The performance comparisons between the context-agnostic and the proposed contextual models are also notable in Fig. 7. A context-agnostic model is an implementation of the same kind of regression model trained without the context data. For the 70%-30% evaluation, the contextual pipeline outperforms context-agnostic models for every choice of regression model. To evaluate robustness and generalizability, we conduct this analysis across multiple train-test percentages; the result is presented in Fig. 9. In this figure, downward arrows refer to loss reductions, i.e. performance improvements. Unsurprisingly, the inference performance slightly worsens with the reduction in training data. But, differences between context-agnostic and contextual models remain steady across the spectrum. For light-weight models (not required to store all samples or transformations) like GLM, RF, and SVM, the performance improves more dominantly than for neighborhood-based heavy-weight (need to store the training set) models such as GPR and NCA. Moreover, the impact of context is higher for inferring VE than for BR, as the arrows are longer for Fig. 9(b), highlighting the effect of volumetric variations.

### C. Biomarker Relevance

From the feature importance of the wearable ECG features, we acquire the relevance of the discovered biomarkers, namely R-height, R-width, T-height, T-width, and R-R interval, in inferring the respiratory parameters. Fig. 10 shows the percentage relevance of the biomarkers across different activities. The relative relevance ranges from 14% to 27% for BR and from 13% to 28% for VE. For low intensity activities, the biomarker relevance are uniformly distributed; during rest and walk, the average standard deviation is only about 2%. Heart-rate or RR biomarker shows more relevance to BR during rest compared to other activities. For ambulatory high intensity activities, like running and biking, both R- and T-wave heights show higher relevance to BR and VE, totaling about 50%, whereas median relevance for RR remains at <15%. T-width shows large relevance to VE during running.

Relevance of the biomarkers represents how well the related features capture the variance of BR or VE, which is also proportional to the entropy and variance within the biomarker clusters. Such relevance also indicates variance along the related electro-cardiac functionality. Thus, the above analysis can be used to interpret the inference process by highlighting the contextual variations in the physiological functions. For example, during rest, the variation in the heart rhythm appears as a leading factor of the cardio-respiratory coupling, captured by the BR inference models. Similarly, the models indicate how high exertion associated activities disturb cardio-respiratory coupling time and impact the ECG wave morphologies. Related studies in cardio-respiratory coupling and exercise stress tests have identified similar variations in the ECG and the heart functions across physical exercises [59-61]. Respiratory coupling along those contexts are still under active research, and this work adds the wearable biomarkers to that arsenal.

### VIII. CONCLUSION

Context matters; specially in addressing confusion and uncertainties in real-world scenarios. This paper presents a novel contextual inference pipeline for inferring respiratory parameters from wearable sensor signals. We implement two pipelines for estimating breathing rate and minute ventilation. They outperform state-of-the-art solutions and achieve mean absolute error of only 1.17 Br/min for BR and 1.39 L/min for VE. We also evaluate the generalizability and robustness by reducing the train-test data size, even down to 20%-80%. Moreover, we interpret the model predictions by identifying wearable ECG biomarkers and using their predictive relevance.

Further improvements of the proposed pipeline can be achieved by parametrizing the contextual aggregator and learning it from data. Similarly, the biomarker relevance aggregation can be made data-driven rather than rule-based. These changes require more ground truth data and physiological explanation from clinical experts. Moreover, the metric and error margin for such inference to be useful in preventive intervention design need to be investigated further in inter-disciplinary studies.

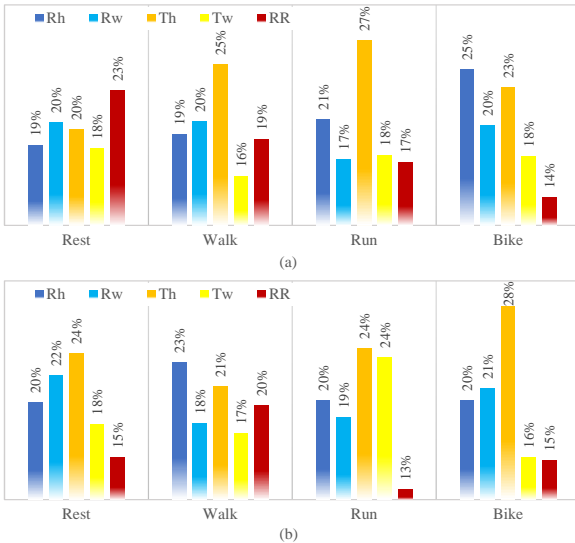

Fig. 10. Relevance of ECG biomarkers to (a) breathing rate and (b) minute ventilation during different physical activities.